\documentclass[graybox]{svmult}

\usepackage{type1cm}        
\usepackage{makeidx}         
\usepackage{graphicx}        
\usepackage{multicol}        
\usepackage{multirow}
\usepackage[percent]{overpic} 
\usepackage[bottom]{footmisc}
\usepackage{tikz} 
\usetikzlibrary{shapes.geometric, arrows}
\usepackage{pgfplots} 
\pgfplotsset{width=\linewidth, compat=1.6} 
\usepackage{pgfplotstable} 

\usepackage{newtxtext}       %
\usepackage{newtxmath}       

\usepackage{booktabs}

\usepackage[normalem]{ulem} 

\makeindex             

\graphicspath{figures}
\begin{document}

\title*{DNS of the Early Phase of Oblique Droplet Impact on Thin Films with FS3D}
\author{Jonathan Lukas Stober, Johanna Potyka, Matthias Ibach, Bernhard Weigand and Kathrin Schulte}
\authorrunning{J.L. Stober, J. Potyka, M. Ibach, B. Weigand and K. Schulte}
\institute{Jonathan Stober and Kathrin Schulte \at Institute of Aerospace Thermodynamics (ITLR), University of Stuttgart, Pfaffenwaldring 31, 70569~Stuttgart, Germany, \email{jonathan.stober@itlr.uni-stuttgart.de} and \email{kathrin.schulte@itlr.uni-stuttgart.de}} 
%
%
\maketitle

\abstract{
Spray impacts occur in several environmental and technical applications.
The impact of droplets at different angles onto walls covered with a thin film of the same liquid can be regarded as an elementary process here. 
Direct Numerical Simulations (DNS) provide an important contribution to the understanding and modelling of the impact outcome, which might be associated with the formation of a crown and ejection of secondary droplets. Thus, we gain detailed information about, e.g.\ the flow field and shape of the interface, which are not accessible in experiments.
This chapter presents a DNS study of the early crown formation mechanisms present at an oblique droplet impact on a thin film, as well as a grid study showing the resolution required to resolve the impact's details.
Highly resolved simulations in large domains require continuous development of the numerical solver's efficiency.
The performance of different cycles of the multigrid (MG) solver for the solution of the pressure Poisson equation was compared, and a F-cycle was added.
Furthermore, we implemented a hybrid MPI + OpenMP parallelisation, which both increases the scaling limit further. 
Additionally, studies on the strong and weak scaling are conducted.
The choice of the F- and V- cycle in the MG-solver and the additional hybrid parallelisation increased the achieved computed cycles per hour (CPH) by a factor of~$12.4$ compared to the formerly employed setup. %
The nodes efficiently usable were increased by a factor of $16$. Both, the close to linear scaling regime for the strong scaling and the almost constant performance regime for the weak scaling were increased by this factor.

}

\section{Introduction}
\label{sec:Introduction}
Sprays occur in various environmental and technical applications, such as soil erosion, medical sprays and cleaning processes, with many droplets impacting at different angles and velocities onto solid walls covered with a film of the same liquid. The thickness of such films ranges from thin films to deep pools depending on the application. %
The impact of a single droplet might result in the formation of a crown and the ejection of secondary droplets. %
The comprehensive physical description of this elementary process is of great importance to optimise such applications. %
Many studies focus on the development of the crown shape as well as the differentiation between the impact regimes \textit{deposition} and \textit{splashing}. %
The latter results in the detachment of secondary droplets from the unstable rim of a crown. %
Most investigations focus on a normal droplet impact, i.e.\ with 90$^\circ$, onto the wall film. %
Cossali et al. \cite{Cossali1997} conducted a large experimental study and derived an empirical correlation, which describes the deposition/splashing threshold as a function of the fluid properties, impact conditions and film thickness. %
The derivation of empirical thresholds for a large variation of impact conditions was also the aim of other studies \cite{VanderWal2005}, as well as the characterisation of the morphology after the impact \cite{Geppert2017,Rioboo2003}. %
Yarin and Weiss \cite{Yarin1995} on the other hand, described the formation of the crown analytically with a kinematic discontinuity and derived a model predicting its expansion. %
Roisman and Tropea \cite{Roisman2002} extended this concept to oblique impacts by allowing arbitrary velocity fields in the wall film after impact. %
Their model predicts the velocity of the upward rising sheet, crown shape and height. %
Numerical studies on the crown height were conducted with different simulation approaches such as Lattice-Boltzmann method \cite{Cheng2014}, 
the Coupled Level set and Volume of Fluid method (CLSVOF) \cite{Liang2014b}, %
or the  Volume of Fluid (VOF) method, which is applied in our code Free Surface 3D (FS3D) \cite{Rieber1999,Kaufmann2018}. %

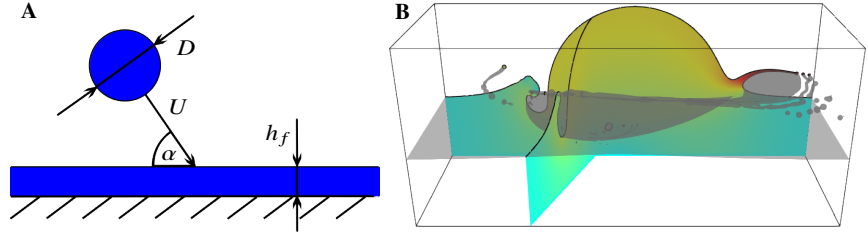
\begin{figure}[ht]
	\begin{center}
		\begin{overpic}[width=0.42\columnwidth]{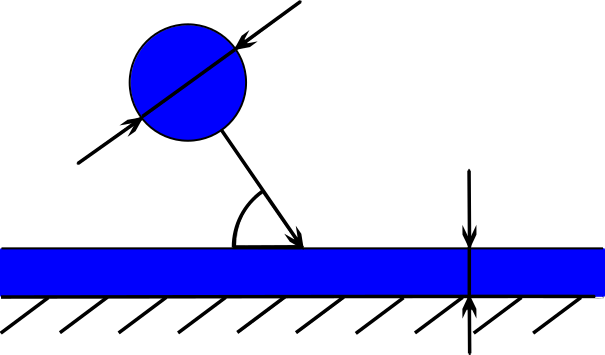}
			\put (3, 58) {\textbf{A}}
			\put (45,48) {\small$D$}
			\put (40.5,19.5) {\small$\alpha$}
			\put (43,32) {\small$U$}
			\put (69,25) {\small$h_f$}
		\end{overpic}
		\begin{overpic}[width=0.55\columnwidth]{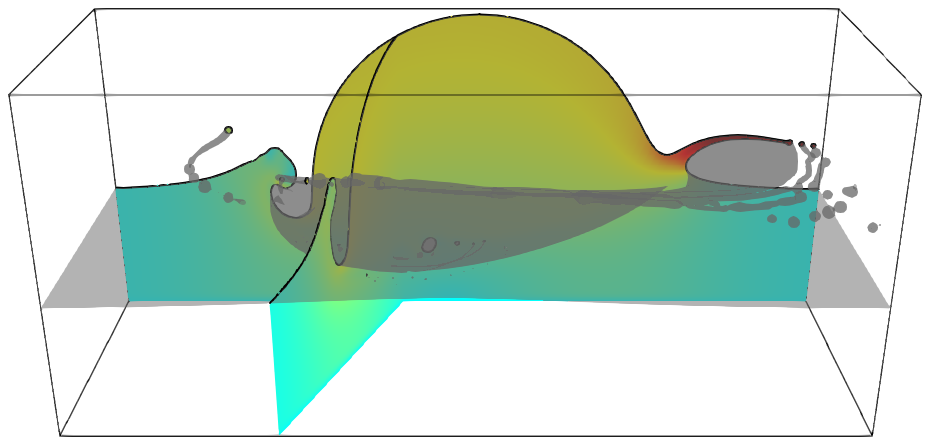}
			\put (2, 44) {\textbf{B}}
		\end{overpic}
		\vspace{1mm}
		\caption{A) Sketch of an oblique droplet impact on a wall film, B) 3D visualisation of the early phase ($\tau=tU/D=$~0.5) of impact.}
		\label{fig:Sketch}
	\end{center}
\end{figure}

In real applications, droplets impact not normally to the wall but at an angle $\alpha$, see the sketch in Figure~\ref{fig:Sketch}~A, which results in different crown shapes and splashing conditions \cite{Okawa2008}. %
Brambilla et al. \cite{Brambilla2015} detected different asymmetric crown shapes in their numerical study with impact angles ranging from 10$^\circ < \alpha <$~90$^\circ$ and a thin wall film, $h_f/D=$~0.116 with $h_f$ as the film height and $D$ the droplet diameter. %
Chen et al. \cite{Chen2020} investigated the onset of splashing in their numerical study on thin films for different impact angles, 45$^\circ < \alpha <$~75$^\circ$. %
Their results confirm an influence of the angle on the crown shape. %
However, no detailed investigation was carried out to compare and classify the mechanisms forming the crown in the early times after the impact, $\tau=tU/D<$~0.5, with t being the time and U the droplet's velocity. %
This early phase is of great importance, as it determines the further development of the impact morphology, but, on the other hand, it is difficult to investigate experimentally, because of its small temporal and spatial scales. %
Furthermore, information on the internal flow structure are required for a better understanding and prediction of the crown formation. %
Therefore, highly resolved Direct Numerical Simulations (DNS) are necessary, see an example in Figure~\ref{fig:Sketch}~B, that allow us to identify two
different crown formation mechanisms on both sides of the crown, explained in detail in Sec.~\ref{sec:results}.  
The simulation of the early crown formation is challenging due to small time scales, high velocities and the small size of the developing liquid structures. %
This requires high spatial and temporal resolutions, which can only be achieved with an efficient parallelisation of the code and the use of high performance clusters. %

First, this work provides an overview of our in-house multi-phase simulation code Free Surface 3D (FS3D), see Sec.~\ref{sec:methods}. %
After that, the computational setup for the simulation of an oblique droplet impact onto a wall film is presented, see Sec.~\ref{sec:setup}. %
The simulation results are shown prior to the grid study to explain the different mechanisms of crown formation. %
The resolution required for reproducing relevant details of the impact is determined by comparing the shape of the interface.  %
Recent performance enhancements are presented in the second part of this report, see Sec.~\ref{sec:Com_Perf}, which enabled the study of such large and complex interactions by increasing the scaling limit. %
This includes a description of the different multi-grid cycles and the hybrid MPI + OpenMP parallelisation as well as the results of a strong and weak scaling. %

\section{Mathematical Description and Numerical Approach}  \label{sec:methods}

The ITLR's in-house program package FS3D was developed since the 1990s for the direct numerical simulation of highly dynamic multiphase problems like droplet deformation \cite{Reutzsch2019a}, grouping \cite{Ibach2022}, single and multicomponent droplet collisions \cite{Potyka2022}, atomisation of liquid jets \cite{Reutzsch2019} as well as droplet impacts onto structured dry and wetted surfaces \cite{Ren2020,Kaufmann2018}. %
To simulate such processes, a high spatial and temporal resolution as well as large computational domains are necessary.
Therefore, the code is parallelised with MPI and OpenMP, and it is continuously optimised to increase its parallel efficiency and scaling on high performance clusters. %

FS3D solves the incompressible Navier-Stokes equations for mass and momentum conservation
\begin{equation}
	\label{eq:mass}
	\partial_t \rho + \nabla \cdot \left( \rho \vec{u} \right) = 0,	
\end{equation}
\begin{equation}
	\label{eq:mom}
	\partial_t \left(  \rho \vec{u}\right)+ \nabla \cdot \left( \rho \vec{u} \otimes \vec{u} \right) = \nabla \cdot \left( \tens{S} - \tens{I} p \right) + \rho \vec{g} + \vec{f}_{\gamma}
\end{equation}
on finite volumes. 
In equations (\ref{eq:mass}) and (\ref{eq:mom}), $\vec{u}$ denotes the velocity vector, $\rho$ the mass density, $p$ the static pressure, $\vec{g}$ the gravitational acceleration, $\tens{S}$ the shear stress tensor and $\tens{I}$ the identity matrix. %
The surface tension forces are introduced as a body force, $\vec{f}_{\gamma}$, in cells close to the phase interface. %
Different surface tension models are implemented in FS3D, such as the continuous surface stress (CSS) model by Lafaurie et al.~\cite{Lafaurie1994} applied in this study. %
An one-field formulation allows to solve one set of conservation equations for all phases with varying physical fluid properties across the interface. %
The VOF method by Hirt and Nichols \cite{Hirt1981} is applied to distinguish between the different phases by introducing the scalar indicator variable,~$f$, 
\begin{equation}
	\label{eq:fdefinition}
	f(\textbf{x},t) = \left\{
	\begin{array}{ll}
		0 & \text{outside the liquid phase},\\
		(0,1) & \text{at the interface},\\
		1 & \text{inside the liquid phase}
	\end{array}\right.
\end{equation}
that represents the volume fraction of the liquid in each cell. %
A transport equation for $f$ needs to be solved in order to obtain the phases' distribution over time, according to
\begin{equation}
	\label{eq:ftransport}
	\partial_t f + \nabla \cdot \left( f \textbf{u} \right) = 0.
\end{equation}
To increase the accuracy of the corresponding $f$-fluxes, FS3D uses the Piecewise Linear Interface Calculation (PLIC) method by Rider and Kothe \cite{Rider1998}, which calculates the orientation and position of the interface in each cell.
The local physical properties $\psi(\textbf{x},t)$ in each cell are defined by the $f$-variable and the fluid properties of the liquid and gaseous phase, $\psi_{l}$ and $\psi_{g}$ according to
\begin{equation}
	\label{eq:density}
	\psi(\textbf{x},t)= \psi_{l} f(\textbf{x},t) + \psi_{g} \left(1 - f(\textbf{x},t) \right).
\end{equation}
Further details on the numerical implementations and applications of FS3D are given in Eisenschmidt et al. \cite{OverviewFS3D2016}.
A group of eleven researchers is currently developing and using this code.
Within the report period from April 2022 until March 2023, approximately 45 million CPU core-hours were spent in total.

\section{Simulation Results}

First, this section presents the computational setup. %
After that, prior to the grid study, the simulation results are presented and the process of crown formation is explained. %
The subsequent comparison of the shape of the interface for different grid resolutions determines the required resolution necessary to reproduce relevant details of the impact. %

\subsection{Computational Setup}
\label{sec:setup}
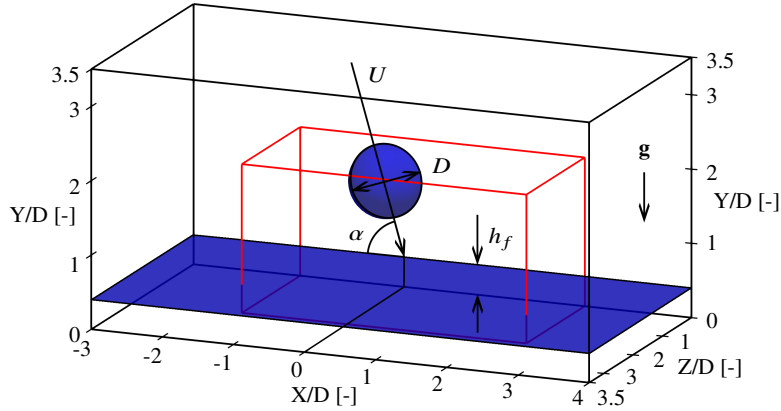
\begin{figure}[ht]
	\begin{center}
		\begin{overpic}[width=0.8\columnwidth]{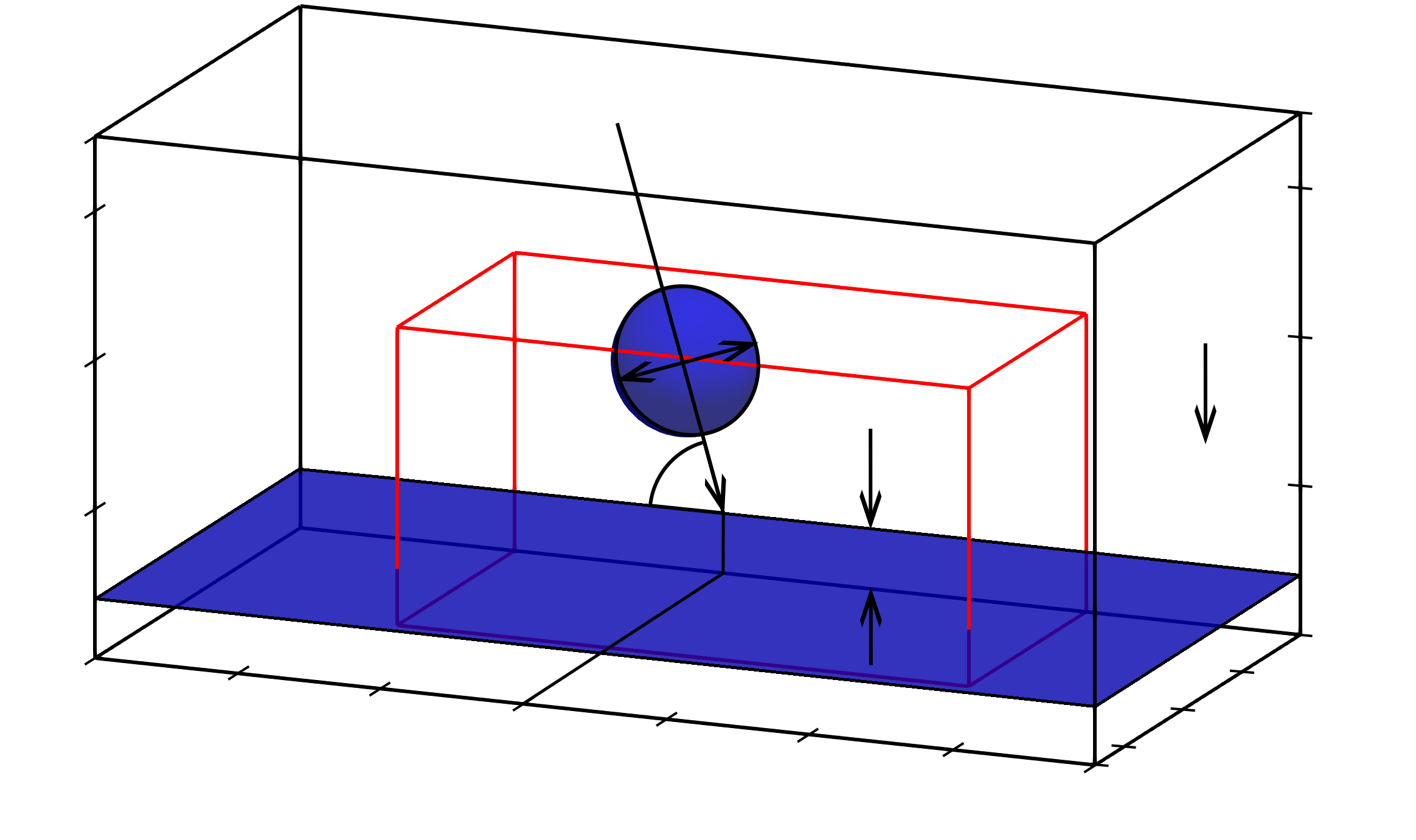}
            \put (55,35) {$D$}
			\put (43,25.5) {$\alpha$}
			\put (46,48) {$U$}
			\put (63,25) {$h_f$}
			\put (84.4,37.5) {$\vec{g}$}
			\put (35,2.5) {X/D [-]}
			\put (-5,28) {Y/D [-]}
			\put (95,30) {Y/D [-]}
			\put (90,6.5) {Z/D [-]}
			\put (3.5,11) {0}
			\put (3.5,21) {1}
			\put (3.5,32) {2}
			\put (3.5,43) {3}
			\put (1,48) {3.5}
			\put (94,13) {0}
			\put (94,24) {1}
			\put (94,34) {2}
			\put (94,45) {3}
			\put (94,50) {3.5}
			\put (90.2,10) {1}
			\put (86.5,7.5) {2}
			\put (82.6,5) {3}
			\put (78.5,2) {3.5}
			\put (4,9) {-3}
			\put (14,8) {-2}
			\put (24,7) {-1}
			\put (35,6) {0}
			\put (45,5) {1}
			\put (55,4) {2}
			\put (65,3) {3}
			\put (75,2) {4}
		\end{overpic}
		\caption{Sketch of the computational domain showing the initial condition and the coordinate system for grid I-IV. The droplet impacts at $X/D=$~0.}
		\label{fig:CompDomain}
		\vspace{-8mm}
	\end{center}
\end{figure}

The previously described program package FS3D is employed to simulate the early phase of crown formation during an oblique droplet impact onto a thin film. 
Figure~\ref{fig:CompDomain} shows a sketch of the impact of a droplet with a diameter $D=$~2.4~mm onto a wall film of the same liquid of height $h_f=$~0.96~mm. %
The impact velocity is $U=$~3.3~m/s and the impact angle is $\alpha=$~60$^\circ$.
A cuboidal computational domain is used and it exploits the symmetry of the impact scenario using a symmetry boundary condition at $Z/D=$~0. 
For the bottom wall, $Y/D=$~0, a no-slip boundary was applied and all other boundaries use zero gradient conditions.
The fluids of our choice are Hexadecane for the droplet and the film and air as ambient gas. The relevant material properties are given in Table~\ref{tab:physprop}. %
The impact conditions are summarised in the following dimensionless parameters, Weber number $We=\rho_l D U^2/\sigma=$~750, Ohnesorge number $Oh=\mu_l/\sqrt{\rho_l D \sigma}=$~0.015 and the non-dimensional film thickness $\delta=h_f/D=$~0.4, which represents the thin film regime. 
The centre of the droplet is initialised 1~D above the wall film.
The initial contact with the wall film happens at $\tau=0$. 
\begin{table}[ht]
	\caption{Physical properties of air and Hexadecane.}
	\label{tab:physprop}
	\centering
	\begin{tabular}{p{1.8cm}p{2.5cm}p{3.6cm}p{3.3cm}}
		\hline\noalign{\smallskip}
		& Density $\rho$ (kg/m$^3$)   & Dynamic viscosity $\mu$ (Pa\,s) & Surface tension $\sigma$ (N/m)\\
		\noalign{\smallskip}\svhline\noalign{\smallskip}
		Air & $1.225$ & $1.82\cdot 10^{-5}$ & \multirow{2}{*}{$27.5\cdot 10^{-3}$} \\
		Hexadecane & $773.3$ & $3.40\cdot 10^{-3}$ & \\
		\noalign{\smallskip}\hline\noalign{\smallskip}
	\end{tabular}
	\vspace{-4mm}
\end{table}

The Cartesian grid consists of equidistant cubic cells within a refined domain marked by the red box in Figure~\ref{fig:CompDomain}. %
Beyond the refinement, the cell size grows gradually to reduce computational cost. %
The total number of cells in x-direction is twice as large as in y- and z-direction. %
Figure~\ref{fig:CompDomain} shows the extension of the domain and of the grid refinement exemplarily for Grids I-IV.
All parameters of the grids investigated in this study are summarised in Table~\ref{tab:GridResolutions}. %
The number of cells increases for Grids I-IV at a constant domain size. %
A further increase of the resolution was realised by a reduction of the domain size for Grid V and VI. On the one hand, computational costs are reduced, on the other hand, the physical evaluation is limited to early time steps where the crown is still in the refined region. %
The resolutions range accordingly from 87.5 cells$/D$ up to 700 cells$/D$.
A domain decomposition with 64$^3$ cells per process was used for all simulations. %
Grid IV and VI were computed with 8192 MPI processes on 128 nodes. Grid IV ran for approximately 33 hours for a simulation time until $\tau=2.0$, Grid VI 50 hours until $\tau=1.0$. %

\begin{table}[ht]
	\caption{Parameters of the different computational grids investigated in the grid study.}
	\label{tab:GridResolutions}
	\centering
	\begin{tabular}{p{0.45cm}p{1.2cm}p{1.2cm}p{1.2cm}p{1.6cm}p{1.2cm}p{1.2cm}p{1.4cm}p{1.3cm}}
		\hline\noalign{\smallskip}
		&\multicolumn{3}{l}{Domain Size [1$/D$]} & \multicolumn{3}{l}{Refinement Domain [1$/D$]} & \multirow{2}{1.2cm}{Number of Cells} & \multirow{2}{1.3cm}{Resolution [cells$/D$]}  \\
		& $X$ & $Y$ & $Z$ & $X$ & $Y$ & $Z$ & & \\
		\noalign{\smallskip}\svhline\noalign{\smallskip}
		I & [-3, 4] & [0, 3.5]& [0, 3.5]& [-1.5, 2.5]& [0, 2.0]& [0, 2.0] & 256$^3$~$\cdot$~2 & 87.5 \\
		II & [-3, 4] & [0, 3.5]& [0, 3.5]& [-1.5, 2.5]& [0, 2.0]& [0, 2.0] & 512$^3$ $\cdot$ 2 & 175 \\
		III & [-3, 4] & [0, 3.5]& [0, 3.5]& [-1.5, 2.5]& [0, 2.0]& [0, 2.0] & 768$^3$ $\cdot$ 2 & 262.5 \\
		IV & [-3, 4] & [0, 3.5]& [0, 3.5]& [-1.5, 2.5]& [0, 2.0]& [0, 2.0] & 1024$^3$ $\cdot$ 2 & 350 \\
		V & [-2, 2.6] & [0, 2.3]& [0, 2.3]& [-1, 1.7]& [0, 1.3]& [0, 1.3] & 1024$^3$ $\cdot$ 2 & 538.5 \\
		VI & [-1.5, 2] & [0, 2]& [0, 2]& [-0.75, 1.25]& [0, 1.1]& [0, 1.1] & 1024$^3$ $\cdot$ 2 & 700 \\
		\noalign{\smallskip}\hline\noalign{\smallskip}
	\end{tabular}
	\vspace{-4mm}
\end{table}
\subsection{Results and Discussion}
\label{sec:results}
\begin{figure}[!ht]
	\begin{center}
		\begin{overpic}[height=0.20\columnwidth]{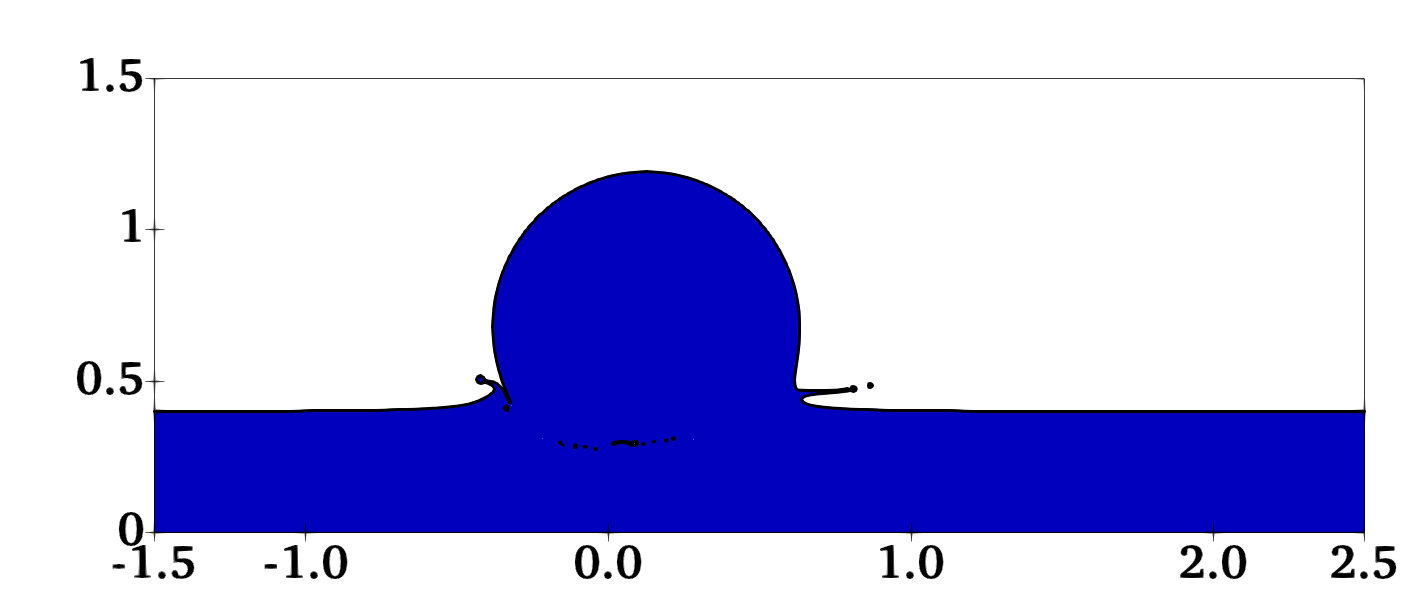}
			\put (14, 30) {\small \textbf{A} \, $\tau=$~0.24}
			\put (50, -2) {\tiny \textbf{X/D [-]}}
			\put (-1, 19.4) {\tiny \textbf{Y/D [-]}}
			\put(40,25){\color{white}\vector(2,-3){6}}
			\put (60,24) {\small impact}
			\put (60,19) {\small direction}
		\end{overpic}
		\begin{overpic}[height=0.20\columnwidth]{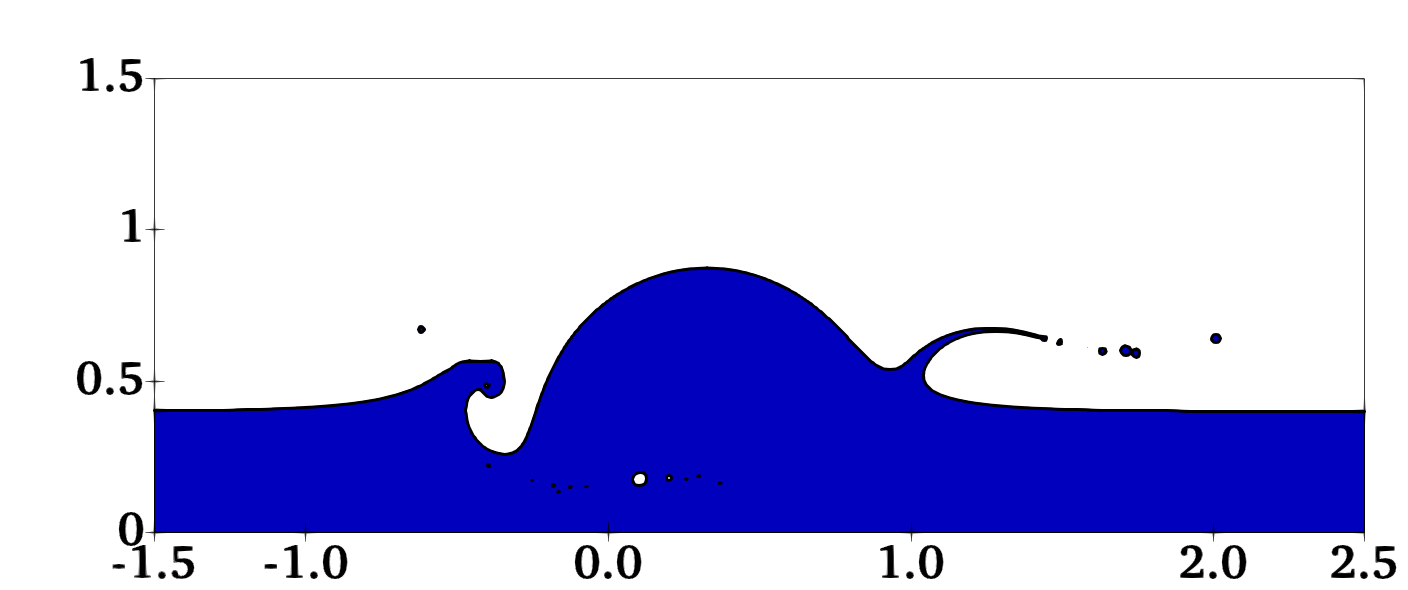}
			\put (14, 30) {\small \textbf{B} \, $\tau=$~0.63}
			\put (50, -2) {\tiny \textbf{X/D [-]}}
			\put (-1, 19.4) {\tiny \textbf{Y/D [-]}}
			\put (62, 21) {\small front}
			\put (17, 21) {\small back}	
		\end{overpic}
		\vspace{1mm}
		\caption{Development of the interface at the symmetry plane for Grid D at two different dimensionless times $\tau~=~tU/D$.}
		\label{fig:MidplaneTotal}
	\end{center}
\end{figure}

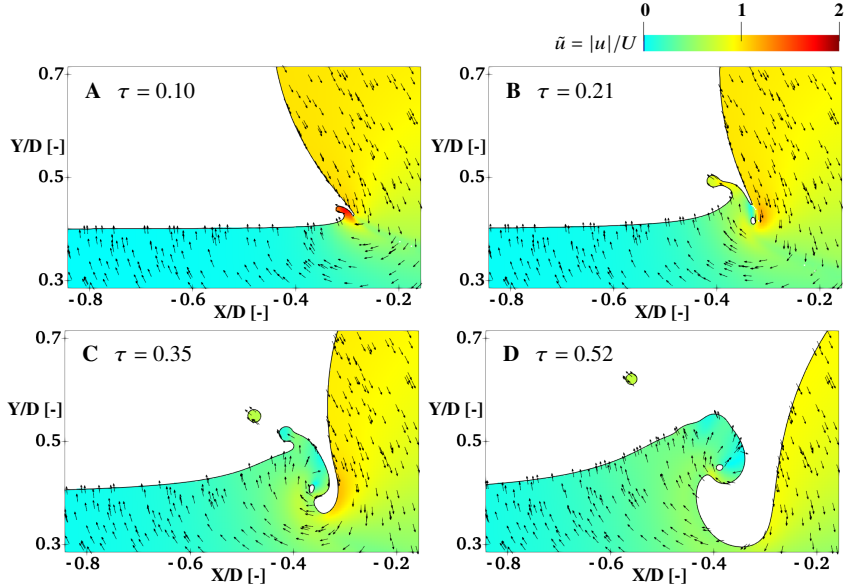
\begin{figure}[!ht]
	\begin{center}
		\hspace{8.15cm}
		\begin{overpic}[width=0.25\columnwidth]{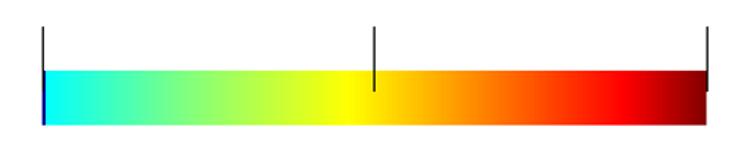}
			\put (-35, 4) {\scriptsize \textbf{$\tilde{u} = |u|/U$}}
			\put (4, 19) {\scriptsize \textbf{0}}
			\put (48, 19) {\scriptsize \textbf{1}}
			\put (92, 19) {\scriptsize \textbf{2}}
		\end{overpic}\\
		\begin{overpic}[height=0.29\columnwidth]{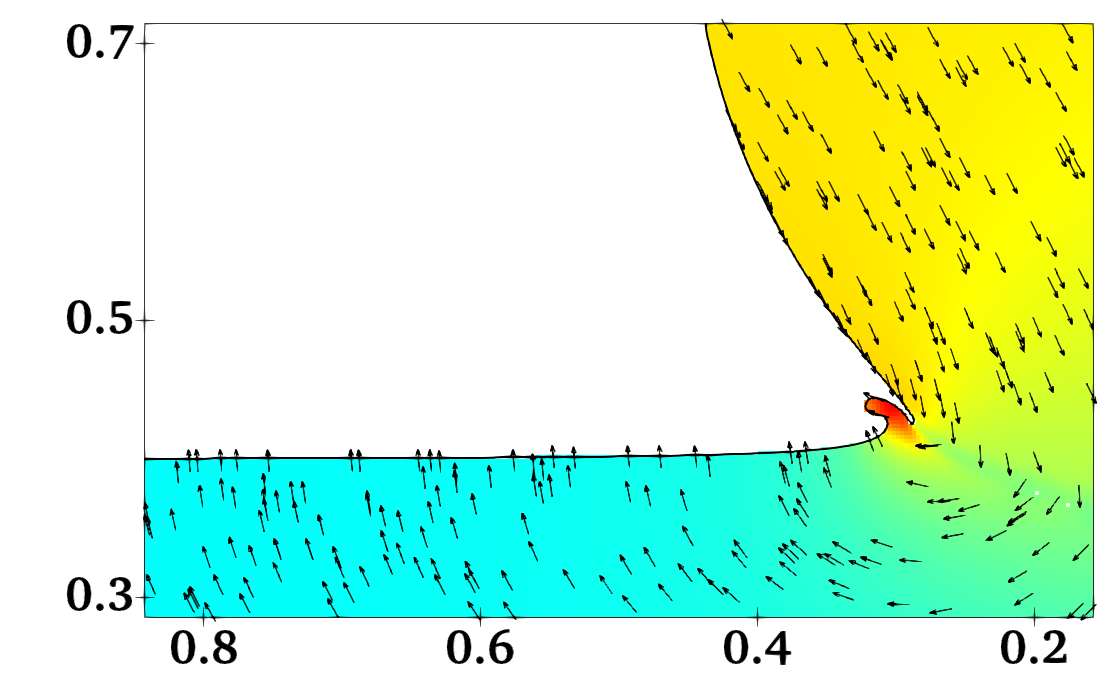}
			\put (17, 52) {\small \textbf{A} \, $\tau=$~0.10}
			\put (48, -1) {\scriptsize \textbf{X/D [-]}}
			\put (-1, 39) {\scriptsize \textbf{Y/D [-]}}
			\put (13, 2.2) {\scriptsize \textbf{-}}
			\put (37.6, 2.2) {\scriptsize \textbf{-}}
			\put (62.3, 2.2) {\scriptsize \textbf{-}}
			\put (87, 2.2) {\scriptsize \textbf{-}}
		\end{overpic}
		\begin{overpic}[height=0.29\columnwidth]{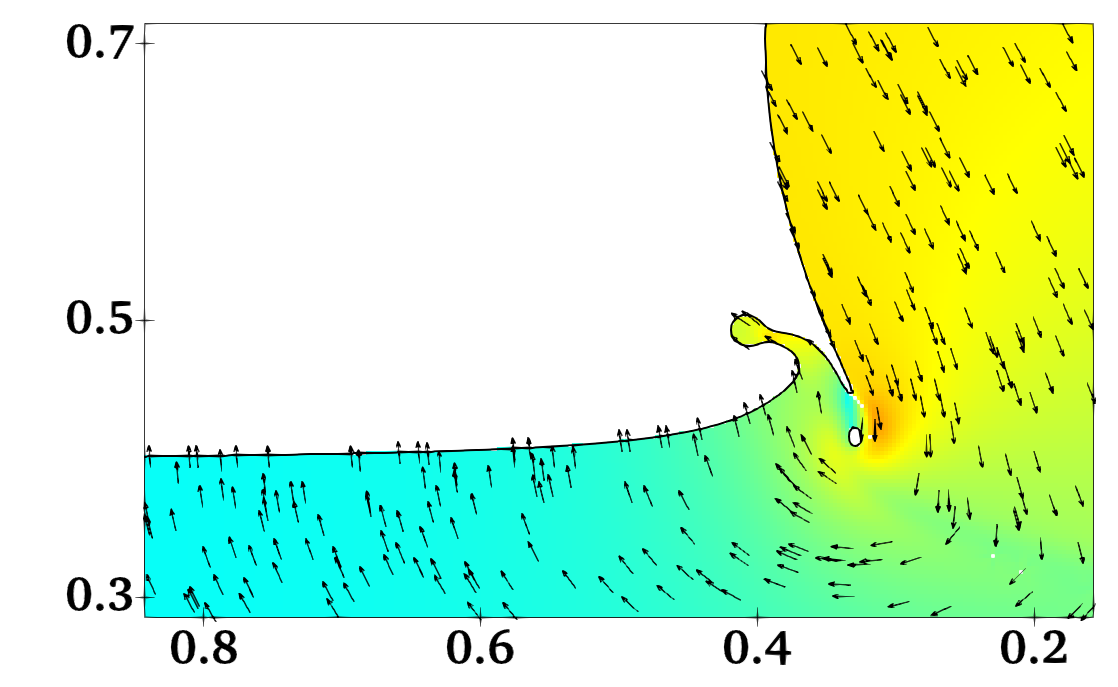}
			\put (17, 52) {\small \textbf{B} \, $\tau=$~0.21}
			\put (48, -1) {\scriptsize \textbf{X/D [-]}}
			\put (-1, 39) {\scriptsize \textbf{Y/D [-]}}
			\put (13, 2.2) {\scriptsize \textbf{-}}
			\put (37.6, 2.2) {\scriptsize \textbf{-}}
			\put (62.3, 2.2) {\scriptsize \textbf{-}}
			\put (87, 2.2) {\scriptsize \textbf{-}}
		\end{overpic}\\
		\vspace{1mm}
		\begin{overpic}[height=0.29\columnwidth]{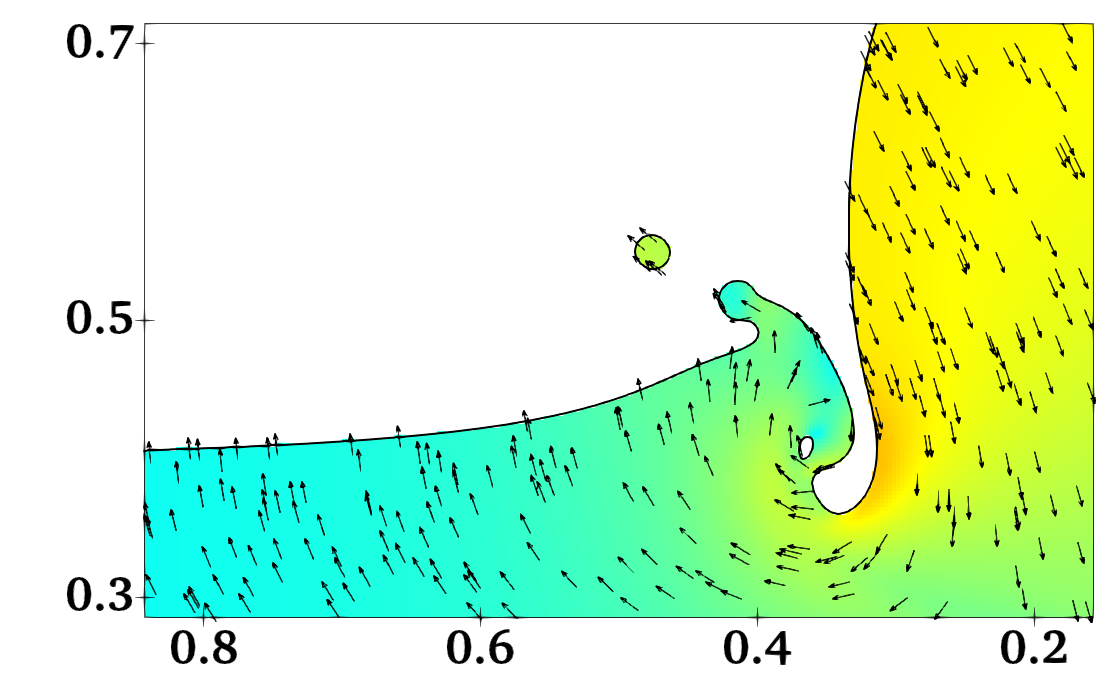}
			\put (17, 52) {\small \textbf{C} \, $\tau=$~0.35}
			\put (48, -1) {\scriptsize \textbf{X/D [-]}}
			\put (-1, 39) {\scriptsize \textbf{Y/D [-]}}
			\put (13, 2.2) {\scriptsize \textbf{-}}
			\put (37.6, 2.2) {\scriptsize \textbf{-}}
			\put (62.3, 2.2) {\scriptsize \textbf{-}}
			\put (87, 2.2) {\scriptsize \textbf{-}}
		\end{overpic}
		\begin{overpic}[height=0.29\columnwidth]{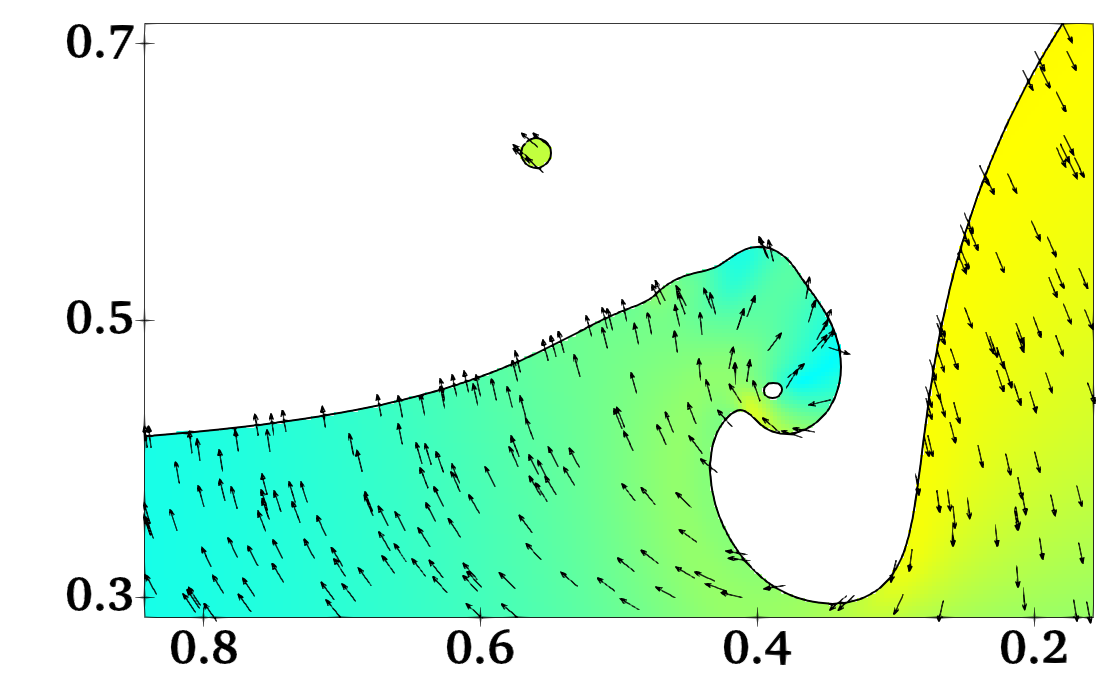}
			\put (17, 52) {\small \textbf{D} \, $\tau=$~0.52}
			\put (48, -1) {\scriptsize \textbf{X/D [-]}}
			\put (-1, 39) {\scriptsize \textbf{Y/D [-]}}
			\put (13, 2.2) {\scriptsize \textbf{-}}
			\put (37.6, 2.2) {\scriptsize \textbf{-}}
			\put (62.3, 2.2) {\scriptsize \textbf{-}}
			\put (87, 2.2) {\scriptsize \textbf{-}}	
		\end{overpic}
		\vspace{1mm}
		\caption{Flow field at the symmetry plane with the dimensionless velocity magnitude $\tilde{u} = |u|/U$ at four moments in time.}
		\label{fig:BacksideVelocity}
	\end{center}
\end{figure}

Figure~\ref{fig:MidplaneTotal} shows the evolution of the fluid interface at the symmetry plane in the early time ($\tau<$~1) after the droplet impact simulated with Grid IV. %
The differences in crown shape on the front, which is thin and bends outwards, compared to the thicker, inward bent crown at the back, are clearly visible. %
At the front, a small lamella forms right after impact and continues to grow outwards and forms the crown.
On the back side, a different mechanism for crown formation is observed. %
Figure~\ref{fig:BacksideVelocity} shows the interface and the velocity field in the fluid. %
The fluid is pushed outwards and forms an initial lamella. %
However, this fluid flow feeding the lamella is pinched-off at $\tau=$~0.21, because it is not able to follow the strong deflection into the lamella due to the horizontal impact velocity component. %
The fluid is then deflected sidewards and upwards with a larger radius, and ambient gas penetrates the wall film.
At $\tau=$~0.35, the initial lamella bends inwards leading to the different crown shape. %
A validation with in-house experiments confirms these two mechanisms on both sides of the crown by comparing the morphology from experiments and simulations. %
The experimental recordings for $\tau=$~[0.19, 0.33, 0.55] confirmed the growth of an initial lamella and its receding as well as resulting in a similar crown shape on the back side. %

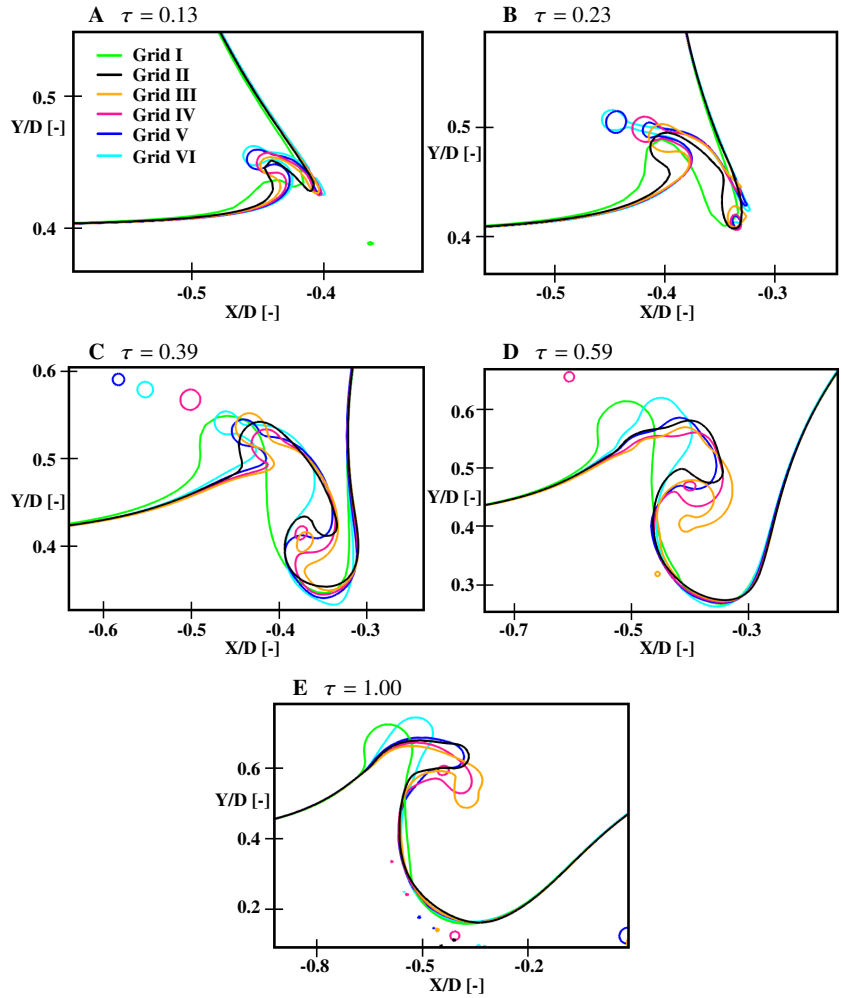
\begin{figure}[t]
	\begin{center}
		\vspace{3mm}
		\quad
		\begin{overpic}[width=0.42\columnwidth]{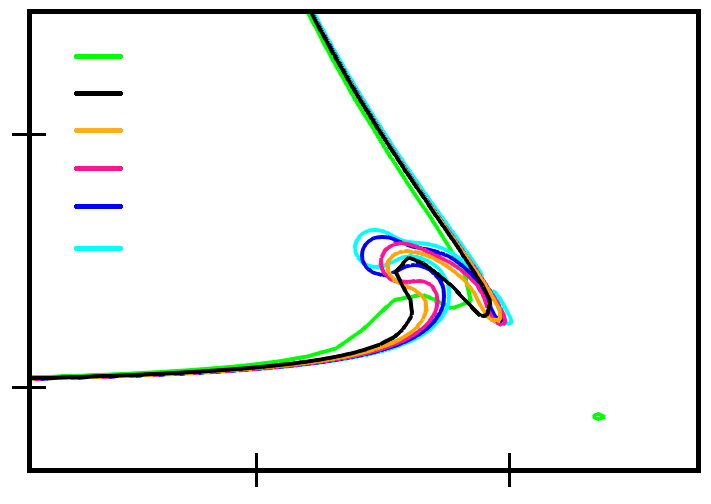}
			\put (8, 72) {\small \textbf{A} \, $\tau=$~0.13}
			\put (45, -8) {\scriptsize \textbf{X/D [-]}}
			\put (-13, 42) {\scriptsize \textbf{Y/D [-]}}
			\put (32, -3) {\scriptsize \textbf{-0.5}}
			\put (67, -3) {\scriptsize \textbf{-0.4}}
			\put (-7, 14.5) {\scriptsize \textbf{0.4}}
			\put (-7, 50) {\scriptsize \textbf{0.5}}
			\put (20, 61.6) {\scriptsize \textbf{Grid I}}
			\put (20, 56.2) {\scriptsize \textbf{Grid II}}
			\put (20, 50.7) {\scriptsize \textbf{Grid III}}
			\put (20, 45.4) {\scriptsize \textbf{Grid IV}}
			\put (20, 39.8) {\scriptsize \textbf{Grid V}}
			\put (20, 34) {\scriptsize \textbf{Grid VI}}
		\end{overpic}\qquad
		\begin{overpic}[width=0.42\columnwidth]{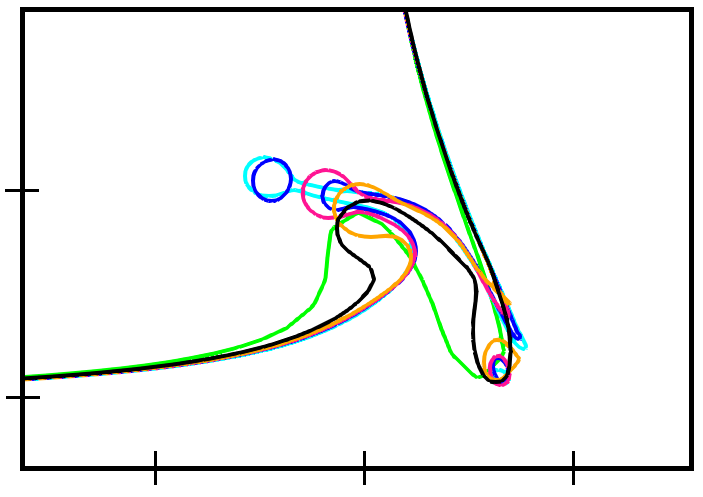}
			\put (8, 72) {\small \textbf{B} \, $\tau=$~0.23}
			\put (45, -8) {\scriptsize \textbf{X/D [-]}}
			\put (-12.5, 35) {\scriptsize \textbf{Y/D [-]}}
			\put (17, -3) {\scriptsize \textbf{-0.5}}
			\put (47, -3) {\scriptsize \textbf{-0.4}}
			\put (77, -3) {\scriptsize \textbf{-0.3}}
			\put (-7, 12) {\scriptsize \textbf{0.4}}
			\put (-7, 42) {\scriptsize \textbf{0.5}}
		\end{overpic}\\
		\vspace{10mm}
		\quad
		\begin{overpic}[width=0.42\columnwidth]{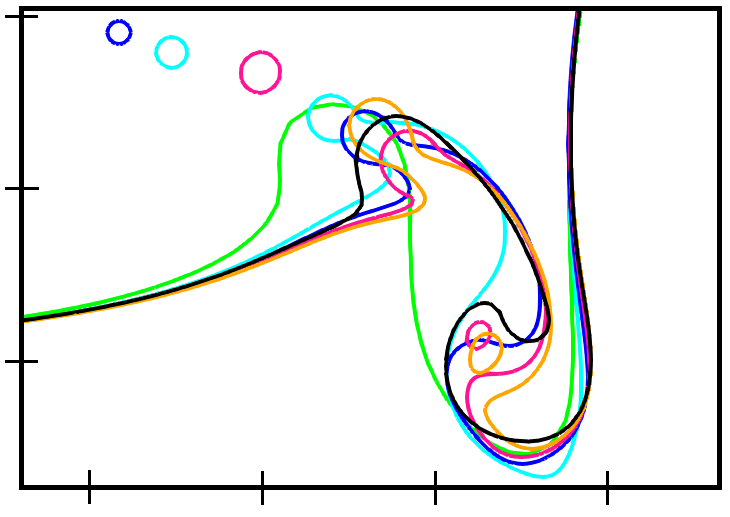}
			\put (8, 72) {\small \textbf{C} \, $\tau=$~0.39}
			\put (45, -8) {\scriptsize \textbf{X/D [-]}}
			\put (-13, 32) {\scriptsize \textbf{Y/D [-]}}
			\put (8, -3) {\scriptsize \textbf{-0.6}}
			\put (32, -3) {\scriptsize \textbf{-0.5}}
			\put (56, -3) {\scriptsize \textbf{-0.4}}
			\put (79, -3) {\scriptsize \textbf{-0.3}}
			\put (-7, 19.5) {\scriptsize \textbf{0.4}}
			\put (-7, 43) {\scriptsize \textbf{0.5}}
			\put (-7, 67) {\scriptsize \textbf{0.6}}
		\end{overpic}\qquad
		\begin{overpic}[width=0.42\columnwidth]{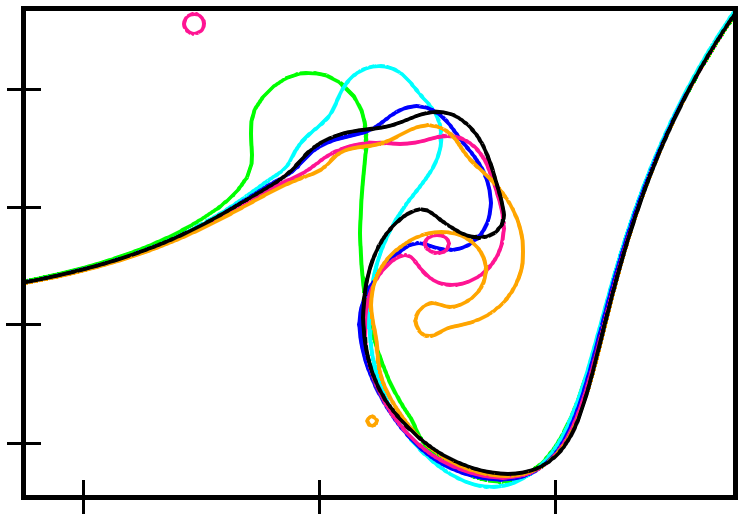}
			\put (8, 72) {\small \textbf{D} \, $\tau=$~0.59}
			\put (45, -8) {\scriptsize \textbf{X/D [-]}}
			\put (-12.5, 33) {\scriptsize \textbf{Y/D [-]}}
			\put (7, -3) {\scriptsize \textbf{-0.7}}
			\put (38, -3) {\scriptsize \textbf{-0.5}}
			\put (70, -3) {\scriptsize \textbf{-0.3}}
			\put (-6, 9) {\scriptsize \textbf{0.3}}
			\put (-6, 24.5) {\scriptsize \textbf{0.4}}
			\put (-6, 40.5) {\scriptsize \textbf{0.5}}
			\put (-6, 56) {\scriptsize \textbf{0.6}}
		\end{overpic}\\
		\vspace{10mm}
		\quad
		\begin{overpic}[width=0.42\columnwidth]{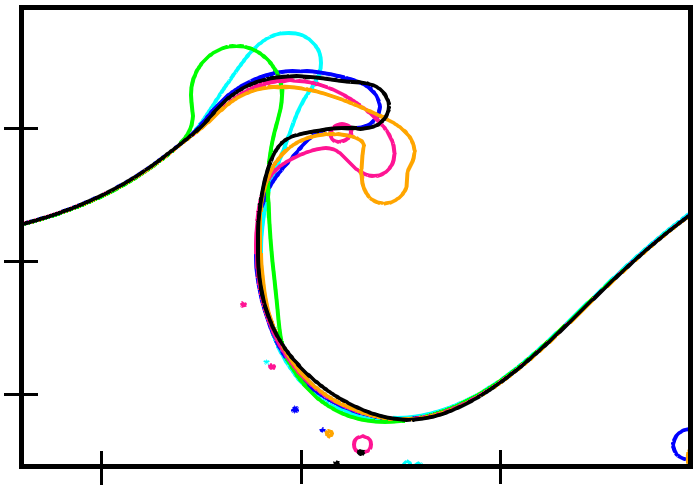}
			\put (8, 72) {\small \textbf{E} \, $\tau=$~1.00}
			\put (45, -8) {\scriptsize \textbf{X/D [-]}}
			\put (-13, 42) {\scriptsize \textbf{Y/D [-]}}
			\put (10, -3) {\scriptsize \textbf{-0.8}}
			\put (38.5, -3) {\scriptsize \textbf{-0.5}}
			\put (68, -3) {\scriptsize \textbf{-0.2}}
			\put (-7, 13) {\scriptsize \textbf{0.2}}
			\put (-7, 31.5) {\scriptsize \textbf{0.4}}
			\put (-7, 50.5) {\scriptsize \textbf{0.6}}
		\end{overpic}
		\vspace{3mm}
		\caption{Interface on the back side of the crown for six different numerical resolutions at five non-dimensional times.}
		\label{fig:GridStudyBackSide}
		\vspace{-7mm}
	\end{center}
\end{figure}

For this highly dynamic and numerically challenging impact phase, a grid study is conducted to investigate the influence of the grid and to identify a sufficient resolution. %
The development of the lamella and the formation of the crown was simulated with six different grid resolutions (Table~\ref{tab:GridResolutions}). %
The results are compared by evaluating the fluid interfaces which are plotted in Figure~\ref{fig:GridStudyBackSide}.
Starting at an early time, $\tau=$~0.13, the lamella is very small and not captured by the lowest resolution (green), see Figure~\ref{fig:GridStudyBackSide} A. %
All other simulations resolve the lamella, but its shape and length differ. %
At $\tau=$~0.23, the growing lamella collides with the bulk of the impacting droplet and leads to an entrapped tunnel of air just below the lamella, which is captured by Grid III (orange) and higher resolutions. %
This tunnel opens up again shortly after, at $\tau=$~0.39, for the simulations with Grid V (blue) and VI (cyan). %
For Grid III (orange), the tunnel opens only at $\tau=$~0.53 which leads to a complete different lamella shape at $\tau=$~0.59 as displayed in Figure~\ref{fig:GridStudyBackSide} D. %
For Grid IV (magenta), the tunnel does not rupture in the investigated time, which leads to entrapped air. %
Comparing the resulting crown shape at $\tau=$~0.59 and $\tau=$~1.00, several finding become apparent. %
First, the simulation results are not yet grid independent. %
Second, the interface of Grid II (black) matches surprisingly well with the second finest resolution (Grid V, blue), while Grid III (orange) deviates the most from the other. %
This can be explained by the fact that the simulation with Grid III is the lowest resolution which captures the collision of the lamella with the droplet. %
However, the exact subsequent process of the tunnel of air and its rupture is not captured correctly. %
Therefore, the results for the later stages are inaccurate. %
The finer resolutions IV (magenta) and V (blue) converge towards the finest VI (cyan). %
In the simulation with Grid II, the collision of the lamella with the droplet is not captured at all. %
Nevertheless, it results in a very similar shape of the crown for later times, $\tau\ge$~0.59. %

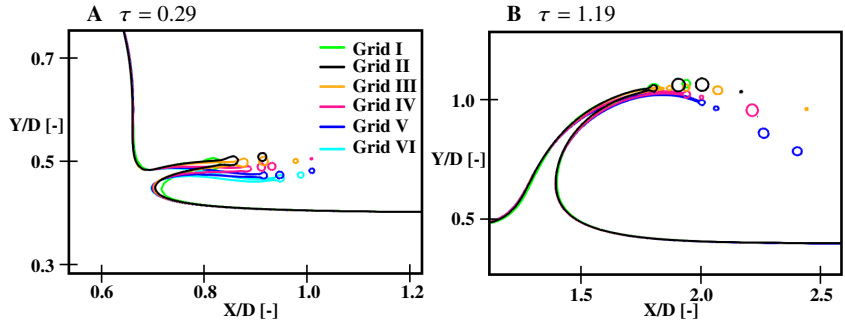
\begin{figure}[t]
	\begin{center}
		\vspace{5mm}
		\quad
		\begin{overpic}[width=0.42\columnwidth]{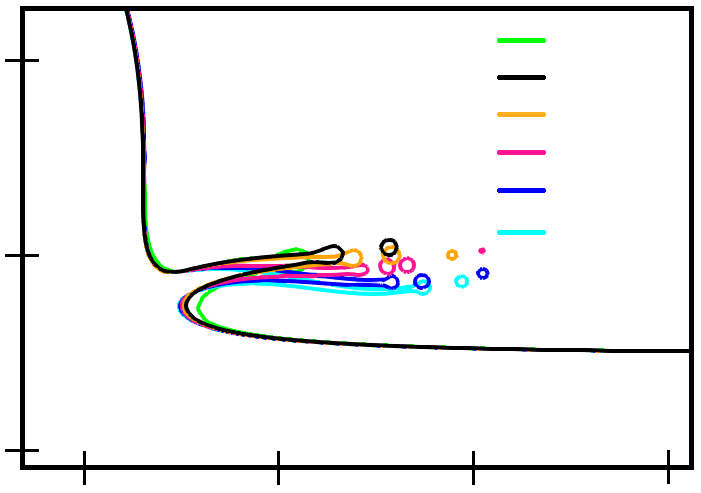}
			\put (8, 72) {\small \textbf{A} \, $\tau=$~0.29}
			\put (45, -8) {\scriptsize \textbf{X/D [-]}}
			\put (-13, 42) {\scriptsize \textbf{Y/D [-]}}
			\put (9, -3) {\scriptsize \textbf{0.6}}
			\put (37, -3) {\scriptsize \textbf{0.8}}
			\put (65, -3) {\scriptsize \textbf{1.0}}
			\put (93, -3) {\scriptsize \textbf{1.2}}
			\put (-7, 4) {\scriptsize \textbf{0.3}}
			\put (-7, 32) {\scriptsize \textbf{0.5}}
			\put (-7, 60) {\scriptsize \textbf{0.7}}
			\put (80, 62.6) {\scriptsize \textbf{Grid I}}
			\put (80, 58.0) {\scriptsize \textbf{Grid II}}
			\put (80, 52.7) {\scriptsize \textbf{Grid III}}
			\put (80, 47.4) {\scriptsize \textbf{Grid IV}}
			\put (80, 41.8) {\scriptsize \textbf{Grid V}}
			\put (80, 36) {\scriptsize \textbf{Grid VI}}
		\end{overpic}
		\qquad
		\begin{overpic}[width=0.42\columnwidth]{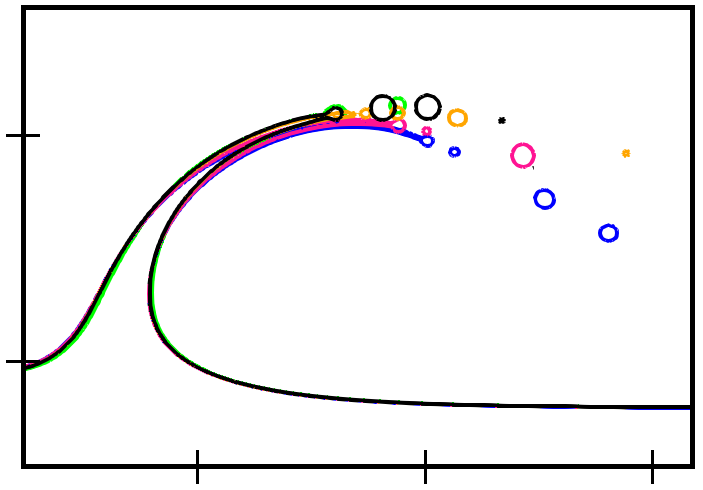}
			\put (8, 72) {\small \textbf{B} \, $\tau=$~1.19}
			\put (45, -8) {\scriptsize \textbf{X/D [-]}}
			\put (-12.5, 33) {\scriptsize \textbf{Y/D [-]}}
			\put (25, -4) {\scriptsize \textbf{1.5}}
			\put (57.5, -4) {\scriptsize \textbf{2.0}}
			\put (90, -4) {\scriptsize \textbf{2.5}}
			\put (-7, 16.5) {\scriptsize \textbf{0.5}}
			\put (-7, 48.5) {\scriptsize \textbf{1.0}}
		\end{overpic}
		\vspace{3mm}
		\caption{Interface on the front side of the crown for six different numerical resolutions at $\tau=$~0.29 (left) and $\tau=$~1.19 (right).}
		\label{fig:GridStudyFrontSide}
		\vspace{-6mm}
	\end{center}
\end{figure}

Figure~\ref{fig:GridStudyFrontSide} shows the interface at the front side of the droplet impact for the different computational grids. %
At the early crown formation phase, see A, the thin lamella is shorter if the resolution is reduced with an unphysical rupture at the tip. %
The height of the lamella decreases for higher grid resolutions. %
The rupture is present also with the highest resolution in our simulations, but not observable in experiments. %
A complete reproduction of the dynamics of the early lamella, including the very thin and fast moving tip, is beyond the possible resolution. %

In Figure~\ref{fig:GridStudyFrontSide}~B, the interface is plotted for the developed crown at $\tau=$1.19. %
The finest grid is not shown for this later time, because the crown has moved outside of the refined grid region. %
The shape of the crown is comparable for all grids. %
But, as in the early stage, a lower resolution leads to an earlier rupture and a shorter crown. %
\\
This grid study reveals that a detailed investigation of the early phase of crown formation after an oblique droplet impact requires high resolutions, and we could not achieve grid convergence for the resolution of all details. %
A grid resolution of $\geq$~350 cells/D is necessary for the reproduction of the lamella at the back side and its collision with the droplet, which justifies the use of Grid IV for the previous investigation. %
The detachment at the tip of the lamella and the rim dynamic remain grid dependent also at later times ($\tau>$~1.0), which influences the size and number of secondary droplets, but the crown shape and the overall behaviour is comparable for different resolutions ($\ge$~350~cells/D). Thus, 
these investigations lay the foundation for further studies about the influence of impact conditions, such as impact angle, impact velocity as well as film thickness and fluid properties, on the crown formation mechanism. %
The simulations allow a detailed view into the process of the droplet impact, which is not accessible from an outside view in the experiments. %
Furthermore, this study revealed the importance of a good scaling efficiency on supercomputers such as the HPE Apollo (Hawk), because a large number of cells is necessary to realise the required high resolutions of important details. %
\section{Computational Performance} \label{sec:Com_Perf}
Constant improvements of FS3D's performance are required to keep up with the supercomputing hardware development and increasingly larger simulation domains and resolutions while maintaining an efficient use of the resources. %
Multiple performance optimisations within FS3D dealt with the most time consuming routines~\cite{HLRSBericht2022} with focus on the multigrid (MG) solver~\cite{HLRSBericht2020}. %
The still most time-consuming MG-solver is worth further optimisation. %
\subsection{Scaling Bottleneck}\label{Sec:IntroBottleneck}
\begin{figure}[!ht]
	\begin{center}
		\begin{overpic}[height=0.24\columnwidth]{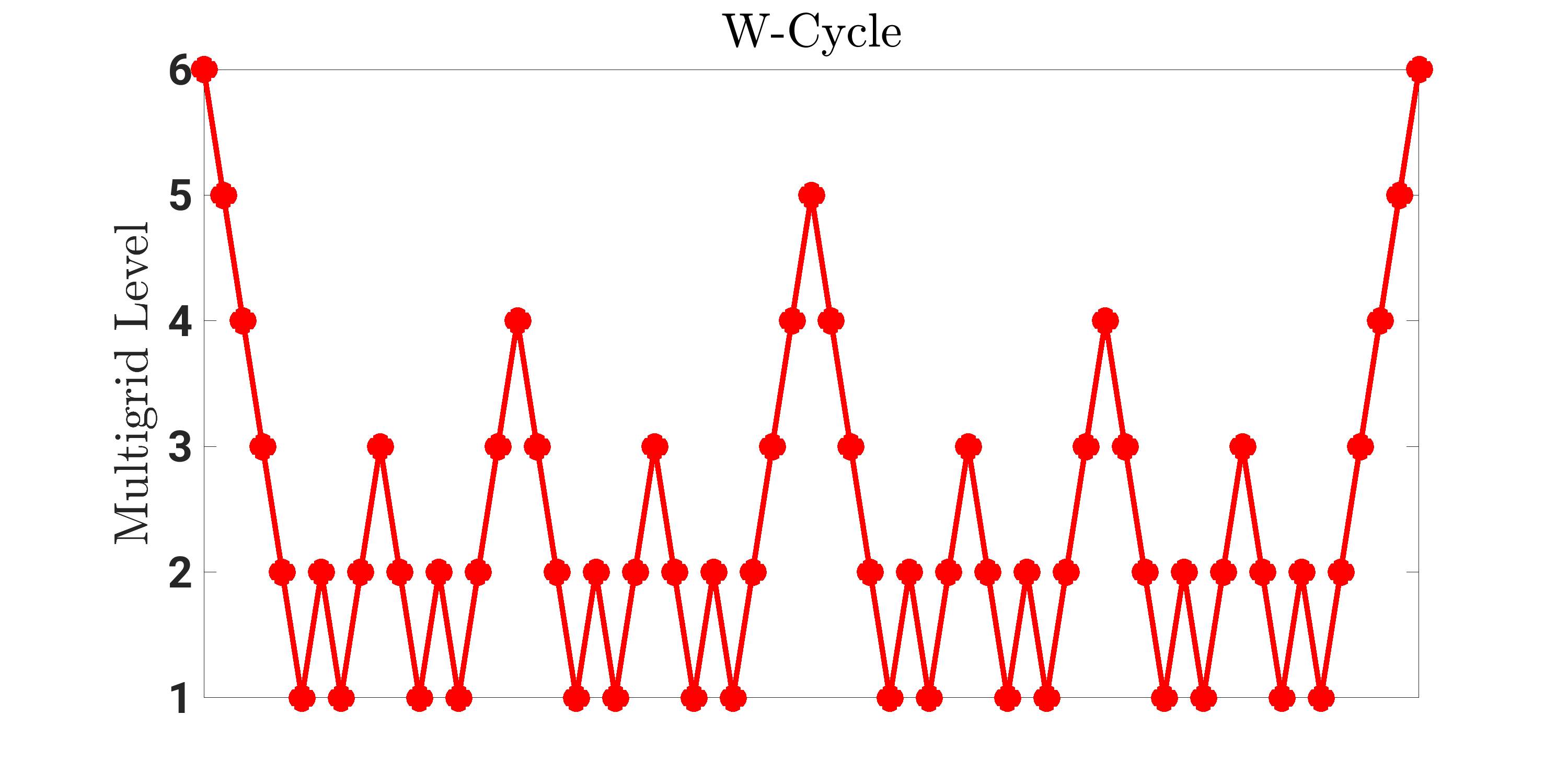}
		\end{overpic}
		\begin{overpic}[height=0.24\columnwidth]{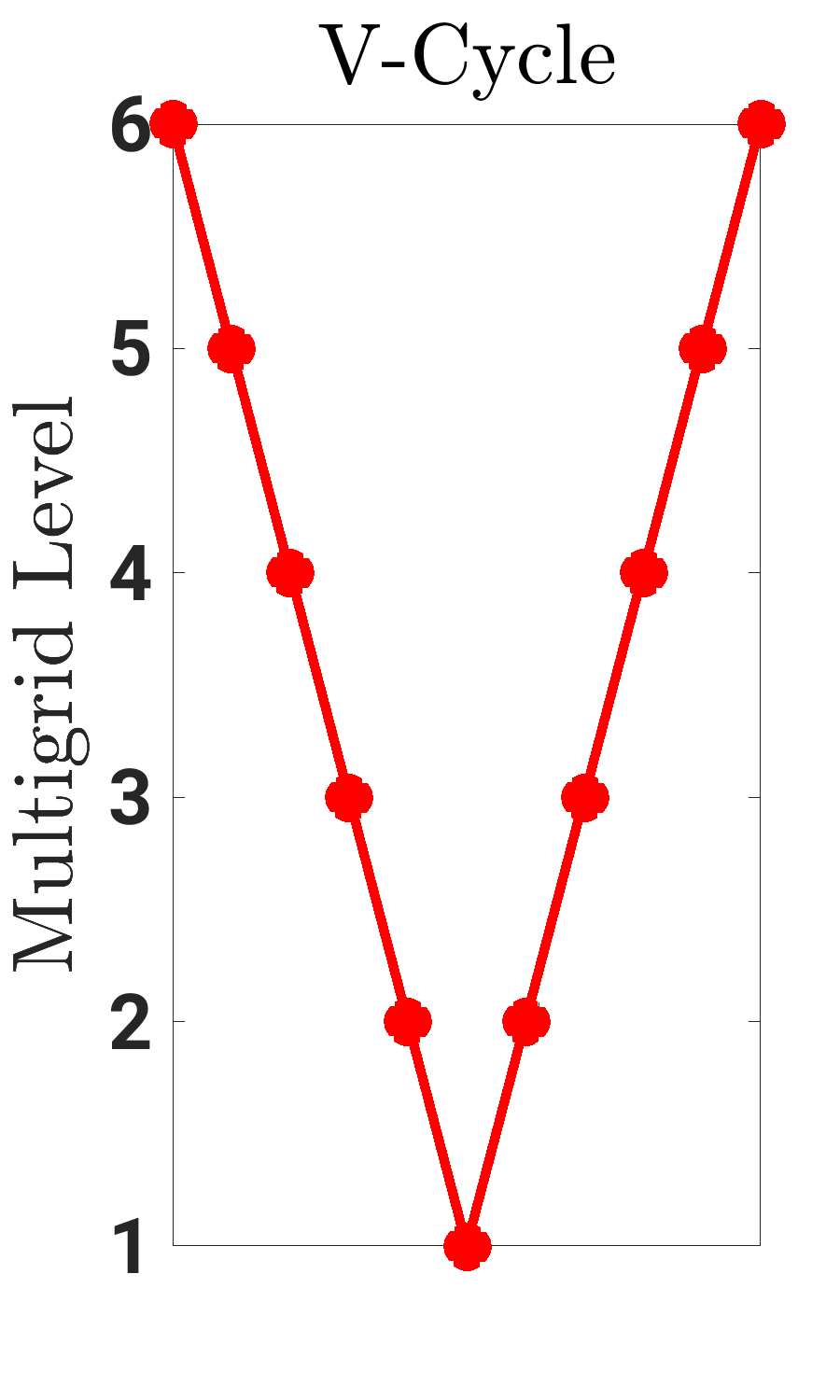}
		\end{overpic}
		\begin{overpic}[height=0.24\columnwidth]{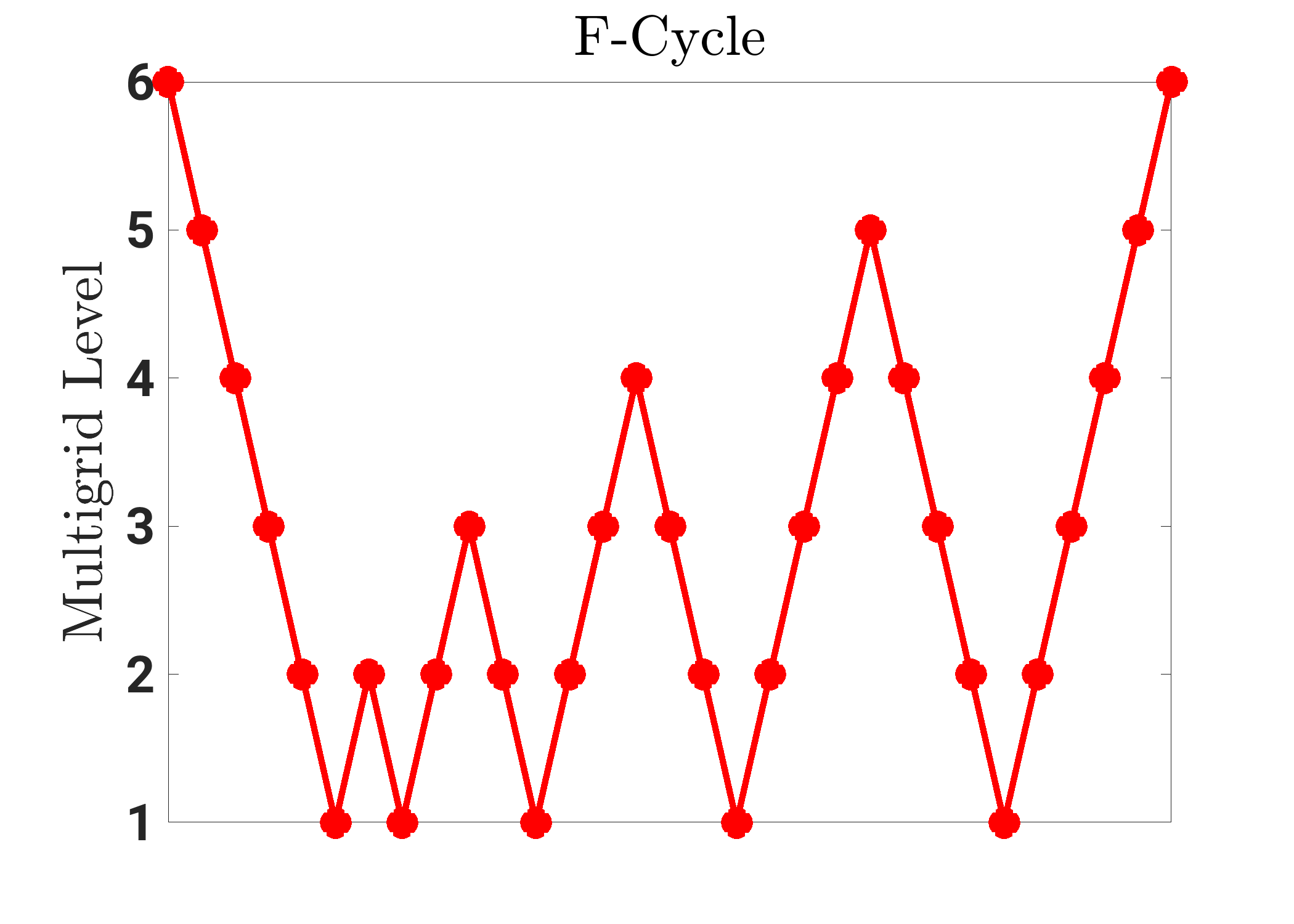}
		\end{overpic}
		\caption{Exemplary representation of the W-, V-, and F-cycle of the MG-solver for six MG-levels. The highest level corresponds to the finest, the lowest to the coarsest grid. The exemplary six levels would be present for $32^3$ cells. The later presented strong scaling example with a domain size of $1024^3$ cells globally has eleven levels in total, the number of levels where restriction and prolongation is executed in parallel decreases with an increasing number of MPI-processes.}
		\label{fig:MGCycles}
	\end{center}
\end{figure}

In FS3D, the numerical solution of the incompressible Navier-Stokes equations necessitates the solution of a Poisson equation for the pressure. %
A MG-method, like the one implemented by Rieber~\cite{Rieber2004} in FS3D, provides means to iteratively solve such huge systems of linear equations efficiently by  solving the problem on consecutively coarser grids leading to faster convergence. %
The next coarser grid is obtained by dividing the number of cells in all three grid dimensions by two, i.e. eight cells are merged.
This leads,	in the extreme case, to a single remaining cell in which the problem can be solved directly.
Within the coarsening, the solution is smoothed and short wave/high frequency errors are reduced. %
In our parallelised code, the restriction is performed on each MPI-process, until no further coarsening is possible anymore. %
After that, all values are collected on one single MPI-process and the restriction is executed again. %
The remaining matrix (1x1x1 for a cuboidal domain decomposition) is solved explicitly, and subsequently the refinement process is started. %
Prolongation steps with analogous smoothing and reduction of long wavelength errors follow until the grid reaches its original resolution.
The coarsening and refinement steps in FS3D were, until now, performed by default with a robust W-cycle as illustrated in Figure~\ref{fig:MGCycles} (left). %
This description already reveals two major problems, leading to the bottleneck in the scaling of FS3D: %
In a strong scaling case, the domain on each process gets smaller.
This leads to a smaller number of restriction steps that can be performed in parallel and the values have to be collected earlier on a single MPI-process. %
The domain size, which has to be solved on a single process thus scales with the number of MPI-processes. 
With each additional level solved on one single MPI-process, the work on this process grows by a factor of two due to the high number of steps in the W-cycle on the high coarsening levels like the Figure~\ref{fig:MGCycles} (left) indicates. 
Furthermore, the increased remaining grid size (scaling with the number of processes involved) adds to the part of the software which is not parallelised. %
Thus, an improved MG-solver should avoid steps on levels solved by one single MPI-process and increase the domain size per MPI-process. %

\subsection{Multigrid Cycle Methods}  
\label{Sec:MGcycles}
The coarsening and refinement steps can be performed in different ways, as illustrated in Figure~\ref{fig:MGCycles}, presenting the three main types of MG-cycles according to Wesseling~\cite{Wesseling1992} with varying trade-offs between the speed of solving one MG-iteration (entire cycle) and the rate of convergence of the solution, namely the V-, W-, and F-cycle.

The simplest one is the V-cycle in which the correction scheme is recursively applied only once at each successive MG-level, while for the W-cycle it is applied twice. %
The V-cycle obviously meets the requirement formulated above that less steps are to be performed on the single MPI-process levels. %
The number of steps in the W-cycle grows by a factor of two, if one MG-level is added, while for the V-cycle only two more steps, one additional restriction and one prolongation, are necessary. %
The drawback of the V-cycle, which is also implemented in FS3D, but was scarcely used until now, is that it converges slower, thus more iterations are necessary. %
Usually, this still leads to a large reduction in steps on the single MPI-process levels, but in cases with high pressure gradients, the V-cycle's convergence might fail due to a lower numerical robustness. %
The W-cycle is very robust, but as the discussion above shows, not well suited to reach a close to linear scaling to a large number of cores. %
Therefore, we now implemented a compromise between V- and W-cycle in FS3D: The F-cycle, which is also a compromise in terms of peak performance as shown later. %
In the cases used for the performance measurements, the averaged number of MG-iterations at peak performance for pure MPI (64 nodes) were $\approx 8.4$ for the W-cycle, $\approx 9.1$ for the F-cycle and $\approx 15.0$ for the V-cycle, but, $15$ iterations of V-cycle result in a shorter simulation time than the $8.4$ iterations in the W-cycle. %
Additional to the reduction of steps, which cannot be solved in parallel, the all-to-one and one-to-all communication can be minimised. %

The introduction of the F-cycle and newly found potential of the V-cycle reduce the bottleneck of serialised computations in the MG-solver for highly parallelised simulations.

\subsection{Hybrid MPI + OpenMP Parallelisation}  
\label{Sec:Hybrid}
As discussed in Sec.~\ref{Sec:IntroBottleneck}, the number of remaining cells, which have to be solved by only one MPI-process, and the costly all-to-one communications govern the scaling limit. %
Increasing the number of cells and thus the domain size per MPI-process is a sensible measure to reduce the fraction of cells which have to be solved on one single core. 
Moreover, the number of processes communicating to one single MPI-process reduces the number of OpenMP threads per MPI-process. %
Implementing a hybrid parallelisation with MPI and additionally using OpenMP solves this problem of reducing MPI-processes but still employing all available cores. %
FS3D was parallelised with OpenMP before to run on vector computers in the past. %
Thus, we enhanced and revised the OpenMP parallelisation by implementing OpenMP on a loop level especially for the recently implemented efficient methods in the most time consuming routines of FS3D \cite{HLRSBericht2022}. %

In addition to moving the levels in the MG-solver solved by one single MPI-process to coarser grids, the load balancing can be improved by a hybrid parallelisation, too. %
Especially on the coarsest levels, which were originally solved by only one process, are now solved by four OpenMP threads.%
Moreover, multiple costly MPI-communications are not required anymore and the shared memory regions reduce the amount of MPI-halos blocking memory in the caches. %
The halos are used for the boundary conditions of the local domain. 
As the memory is shared among OpenMP threads, no boundaries have to be exchanged within a shared memory sub-domain. %
However, one has to consider the limitations the hardware provides, as the performance can worsen significantly instead of the desired improvement: %
Four CPU-cores (CCX) share a common Level 3 (L3) cache on HPE Apollo (Hawk). %
Thus we tested two, four and eight OpenMP threads per MPI-process to investigate the impact of the shared cache. %
This revealed that four OpenMP threads per MPI-process improved the performance compared to using the same number of cores with only MPI-processes, while two or eight OpenMP threads worsened the performance compared to a pure MPI application solving the identical problem. %
Eight OpenMP threads do not share the Level 3 cache, leading to costly memory transfers as no ``first-touch''-principle was implemented. %
Therefore, the following performance analysis employs four OpenMP threads per MPI-process.

\subsection{Strong Scaling} 
\label{sec:Strong_Scaling}
\begin{figure}
\begin{center}
\begin{tikzpicture}
\begin{loglogaxis}[
				axis lines=left,
				width=(\textwidth)*1.0,
				height=(\textwidth)*0.44,
				xmin=3.8,xmax=300,
				ymin=10,ymax=13000,				
				xtick={4,8,16,32,64,128,256},
				xticklabels={4,8,16,32,64,128,256},
            	xlabel={Nodes},
            	ylabel={CPH [1/h]},
            	grid = major,
		        line width=1pt, 
		        legend style={
					at={(0.39,1.2)},
					anchor=north,
					legend columns=2,
					cells={anchor=west},
					rounded corners=2pt,}, 
				]
				\addplot[color=black,dashed,mark=x,mark size = 4pt] table[x=Nodes, y=CPH] {./Txt-files/W-cyclePureMPI.txt};
				\addplot[color=black,mark=x,mark size = 4pt] table[x=Nodes, y=CPH] {./Txt-files/W-cycle_OMP4.txt};
				
				\addplot[color=blue,dashed,mark=o,mark size = 4pt] table[x=Nodes, y=CPH] {./Txt-files/V-cyclePureMPI.txt};
				\addplot[color=blue,mark=o,mark size = 4pt] table[x=Nodes, y=CPH] {./Txt-files/V-cycle_OMP4.txt};
				
				\addplot[color=red,dashed,mark=diamond,mark size = 4pt] table[x=Nodes, y=CPH] {./Txt-files/F-cyclePureMPI.txt};
				\addplot[color=red,mark=diamond,mark size = 4pt] table[x=Nodes, y=CPH] {./Txt-files/F-cycle_OMP4.txt};

\legend{W-cycle pure MPI,
        W-cycle 4 OpenMP Threads,
        V-cycle pure MPI,
        V-cycle 4 OpenMP Threads,        
        F-cycle pure MPI,
        F-cycle 4 OpenMP Threads,        
      }
      
\end{loglogaxis}
\end{tikzpicture}
\begin{tikzpicture}
\begin{loglogaxis}[
				axis lines=left,
				width=(\textwidth)*1.0,
				height=(\textwidth)*0.44,
				xmin=0.007,xmax=400,
				ymin=10,ymax=15000,
                xtick={0.0078125,0.0625,0.5,1,4,8,16,32,64,128,256},
				xticklabels={1/128,8/128,64/128,1,4,8,16,32,64,128,256},
            	xlabel={Nodes},
            	ylabel={CPH [1/h]},
            	grid = major,
		        line width=1pt, 
		        legend style={
					at={(0.39,1.4)},
					anchor=north,
					legend columns=2,
					cells={anchor=west},
					rounded corners=2pt,}, 
				]
				\addplot[color=black,dashed,mark=x,mark size = 4pt] table[x=Nodes, y=CPH] {./Txt-files/W-cyclePureMPI_WS.txt};
				\addplot[color=black,mark=x,mark size = 4pt] table[x=Nodes, y=CPH] {./Txt-files/W-cycle_OMP4_WS.txt};
				
				\addplot[color=blue,dashed,mark=o,mark size = 4pt] table[x=Nodes, y=CPH] {./Txt-files/V-cyclePureMPI_WS.txt};
				\addplot[color=blue,mark=o,mark size = 4pt] table[x=Nodes, y=CPH] {./Txt-files/V-cycle_OMP4_WS.txt};
				
				\addplot[color=red,dashed,mark=diamond,mark size = 4pt] table[x=Nodes, y=CPH] {./Txt-files/F-cyclePureMPI_WS.txt};
				\addplot[color=red,mark=diamond,mark size = 4pt] table[x=Nodes, y=CPH] {./Txt-files/F-cycle_OMP4_WS.txt};

\end{loglogaxis}
\end{tikzpicture}
\end{center}
\vspace{-3mm}
\caption{Strong scaling (top): $1024^3$ cells distributed on an increasing number of nodes. Weak scaling (bottom): $64^3$ cells per core with an increasing number of nodes. Both with and without utilising OpenMP in addition to MPI showing estimated cycles per hour (CPH)} 
\label{Fig:Scaling}
\end{figure}
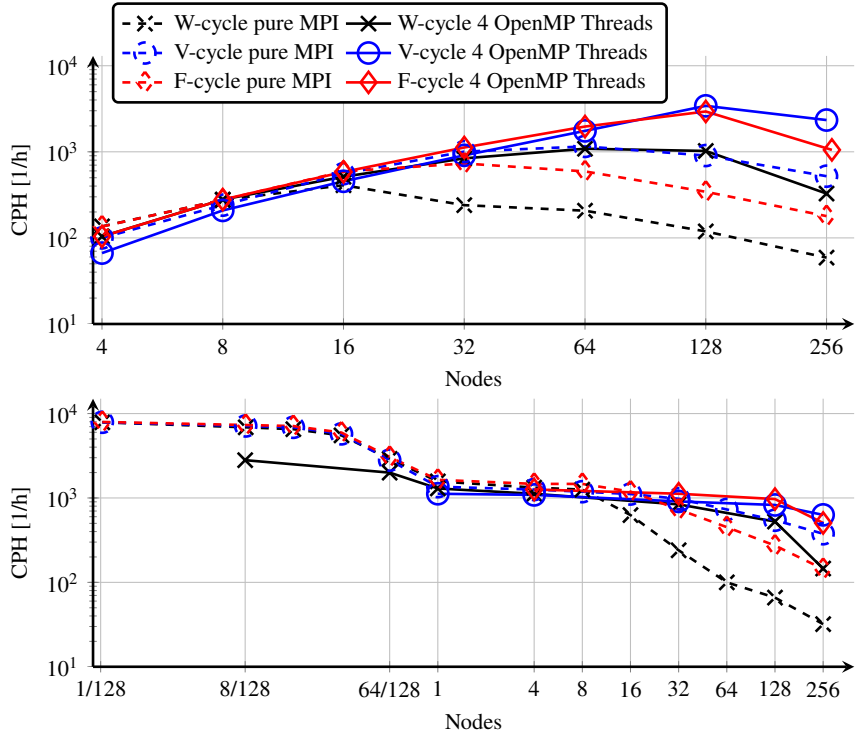
The test case for the performance analysis is an oscillating water droplet within a cuboidal domain. A spatial resolution of $1024^3$ grid cells was used to evaluate the strong scaling of FS3D. %
The strong scaling is shown in Figure~\ref{Fig:Scaling}~(top). %
The figure presents the cycles per hour (CPH) achieved with this test case for the different numbers of nodes used for the parallelisation of the same case. %
The CPH is the number of time steps which can be computed in an hour, extrapolated from a 30~minute run-time measurement. %
As each node on the employed infrastructure consists of $128$ CPU-cores, the pure MPI cases were computed with one MPI-process per core, while the cases with a hybrid parallelisation with four OpenMP threads employed $32$ MPI-processes per node with a stride of four OpenMP threads pinned to a CCX, i.e.\ four CPUs with a shared L3-cache. %
Therefore, we show a comparison of the usage of the same computational resources while solving the same problem size with an identical resolution with different parallelisation techniques and different MG-cycles. %
The number of nodes employed for the parallelisation was increased until the last combination tested showed a decrease in the performance, thus the largest number of nodes tested was $256$ nodes, corresponding to $32^3$ cells per core. %
This is thus a distribution of $32^3$ per MPI-processes for the pure MPI cases and $64\times64\times32$ cells per MPI-process with four OpenMP threads. %
Each case was repeated multiple times, to ensure a constant performance measurement. %

Figure~\ref{Fig:Scaling}~(top) shows that there is not a single optimal choice for the parallelisation technique or MG-cycle method. %
Showing the strong scaling for less than four nodes was not possible with the used test case as the case did not fit into the memory of a single node. %
The W-cycle (pure MPI, dashed black line) serves as the reference case since it was formerly used in most of the conducted simulations shown in previous reports. %
In the following plots, different MG-cycles are denoted in different colours and symbols, pure MPI measurements are shown in dashed lines, hybrid calculations (MPI + OpenMP) in solid lines. %
For the cases where only MPI-parallelisation was used (dashed lines in Figure~\ref{Fig:Scaling}~(top)), the W- and F-cycle outperform the V-cycle in terms of cycles per hour for parallelisation with up to eight nodes. %
The W-, F- and the V-cycle show an almost linear scaling up to $8$, $16$ and $32$ nodes, respectively, before a non-ideal increase in CPH occurs followed by a decrease. %
Just by employing different MG-cycle schemes, a performance boost up to $3$ times the CPH at $32$ nodes for the F-cycle was achieved compared to the reference W-cycle case. The V-cycle even resulted in $5.5$ times the cycles per hour at $64$ nodes at peak performance. %

The additional factor four in the scaling limit is reached by employing the hybrid MPI+OpenMP parallelisation, see the solid lines in Figure~\ref{Fig:Scaling}~(top). %
The F-cycle in combination with the hybrid parallelisation performs best up to $64$ nodes. %
Moreover, the new scaling limit is now $3420$ CPH with the V-cycle, $128$ nodes and a hybrid parallelisation, while the old close to linear scaling limit was at $276$ CPH with the W-cycle, eight nodes and a pure MPI parallelisation. %

\subsection{Weak Scaling} 
\label{sec:Weak_Scaling}

The weak scaling employs the same test case, but with different domain decompositions. 
We kept the amount of grid cells per core constant at $64^3$ and increased the number of used nodes. %
The pure MPI cases were run with $64^3$ cells per MPI-process. With four OpenMP threads per MPI-process for the hybrid cases, the domain size per MPI-process was $128 \times 128 \times 64$. %
Thus, we again compare solving the same problem with the same resources. %

We included the scaling on a single node in the left part of Figure~\ref{Fig:Scaling}~(bottom). %
The weak scaling on a node level (fractions of a node on the Nodes axis) demonstrates that FS3D is memory bandwidth bound: With more than 32 cores (32/128 node) the performance drops.  %
However, the speed-up comparing the performance of $128$ instead of $32$ or $64$ cores is still positive. %
Instead of an ideal factor of four (compared to 32 cores) or two (compared to 64 cores) theoretically achievable, we obtain only a factor of $3.5$ or $1.8$ in the CPH. %
The performance plot also confirms the discussion in Sec.~\ref{Sec:Hybrid} that it does not make sense to employ a hybrid parallelisation on a low number of nodes. %
The problem of a large matrix to solve and many costly all-to-one communications is irrelevant at a low count of MPI-processes, but dominates at larger numbers. %
Beyond the single-node scaling, the performance stays roughly constant again with the pure MPI and F-cycle being the best choice for up to $16$ nodes (2048 cores). %
Adding a hybrid parallelisation with four OpenMP threads at a larger number of nodes further increases the good scaling up to $128$ nodes, again with the F-cycle. %
The V-cycle performs best among the test cases for $256$ nodes, but the performance already drops below the (close to) constant scaling level observed for up to $128$ nodes. %
Thus, the close to constant scaling limit increases by a factor of $16$ in the usable nodes compared to the baseline with the W-cycle and pure MPI.

\section{Conclusion}
The early phase of crown formation during an oblique droplet impact onto a thin wall film was computed with the multiphase DNS code FS3D.
The simulation provides additional information about the liquid gas interface and the internal flow field in the fluid, which can not be derived from experiments.
With the help of the simulations, two different mechanisms of crown formation were observed and could be explained.
The oblique mechanism on the back side of the impact is caused by a pinch-off of the initial lamella due to the horizontal impact velocity component, which prohibits the undisturbed growth of the lamella.
A grid study for this highly dynamic and numerically challenging impact phase revealed that a detailed investigation requires a resolution $\ge$~350~cells/D.
At this resolution, the collision of the lamella with the droplet was adequately resolved.
The detachment at the tip of the lamella remains to be grid dependent, however, the crown shape is resolved sufficiently accurate.

The choice of the F- and V- cycle in the MG-solver and an additional hybrid parallelisation increased the achieved computed cycles per hour (CPH) by a factor of~$\approx 12.4$ for the strong scaling compared to the formerly employed setup at the close to linear scaling limit. %
The strong- and weak scaling was increased to $16$ times more nodes by employing the V- or F-cycle in the MG-solver instead of the W-cycle and introducing a hybrid MPI+OpenMP parallelisation with four OpenMP threads per MPI process.

%


\newenvironment{acknowledgments}%
{\null\begin{center}%
 	\bfseries Acknowledgments\end{center}}%
{\null}
\begin{acknowledgments}
The authors kindly acknowledge the {\it High Performance Computing Center Stuttgart} (HLRS) for support and supply of computational resources on the HPE Apollo (Hawk) platform under the Grant No. FS3D/11142. Additionally, the authors gratefully acknowledge the financial support of the Deutsche Forschungsgemeinschaft (DFG, German Research Foundation) through the project DROPIT/GRK~2160/2 and under Germany's Excellence Strategy - EXC 2075 - 390740016.\\
\end{acknowledgments}

\bibliography{fs3d_HLRS2023report}
\bibliographystyle{spmpsci}


\eject
\end{document}